# Full control of polarization states and phase distributions of light with dual-metasurfaces

Jianxiong Li[1], Shuqi Chen[1*], Haifang Yang[2], Junjie Li[2], Ping Yu[1], Hua Cheng[1], Changzhi Gu[2], and Jianguo Tian[1*]

[1]Laboratory of Weak Light Nonlinear Photonics, Ministry of Education, School of Physics and Teda Applied Physics Institute, Nankai University, Tianjin 300071, China

[2]Beijing National Laboratory for Condensed Matter Physics, Institute of Physics, Chinese Academy of Sciences, P.O.Box 603, Beijing 100190, China

*To whom correspondence should be addressed. E-mail: schen@nankai.edu.cn (S.C.); jjtian@nankai.edu.cn (J.T.)

**Control of the phase and polarization states of light is an important goal for nearly all optical research. The development of an efficient optical component that allows the simultaneous manipulation of the polarization and phase distribution is needed. Traditional methods require the combination of multiple optical devices, and a single optical device cannot easily realize full control of light. We theoretically predict and experimentally verify that our proposed dual-metasurfaces provide an excellent means to simultaneously manipulate the phase and polarization of transmission light at the nanoscale. By introducing a phase gradient along the interface, we achieved a near-perfect anomalous refraction with controllable polarization in the near-infrared region. On the basis of these properties, we created a dual-metasurface capable of generating radially polarized beam, demonstrating the power of full control of light. This work**

**opens exciting avenues toward improving the degrees of freedom in the manipulation of light, including the propagation direction and distribution of the polarization and phase, and may profoundly affect a wide range of plasmonic applications.**

The polarization and the phase of light waves are basic optical properties. The precondition in most optical applications is control and manipulation of the polarization states or phase distributions, which play important roles in the interaction of light with matter (1-3). Birefringence or total internal reflection effects in crystals and polymers are usually used to control the light polarization states; these effects result from the different refractive indices for *s*- and *p*-polarized lights (4). The most common types of modern polarizing prisms are wave-plate and Glan-Taylor prism. The shape of the phase distribution also depends on the variation of the spatial profile of the refractive index or control of the surface topography using lenses, prisms, gratings and holograms. However, these traditional methods face many unresolved issues, such as the large volume of the optical components, a single responding wavelength, and difficulty of miniaturization and integration. In addition, with the ever-expanding field of optics research, the simultaneous manipulation of polarization states and phase distributions has become an urgent need. Traditional methods require the combination of multiple optical devices (5,6), and a single optical device cannot easily realize full control of light.

Metamaterials have enabled the realization of numerous fascinating phenomena and applications, including invisibility cloaking (7,8), a negative index of refraction (9,10), and a metalens beyond the diffraction limit (11). Through the use of specific structural and material designs, the polarization states of light can also be controlled or manipulated at the nanoscale by metamaterials. Polarization conversion devices have been prepared using crossed resonant plasmonic nanostructures (12), metallic cut wire arrays (13), and three-dimensional chiral metamaterials (14). However, full control of the phase distributions of light is not easy with homogeneous metamaterials, which require complex design rules or cannot provide the ability to tailor the phase of the transmitted light from 0 to $2\pi$. Recently, new types of structured plasmonic interfaces that introduce an abrupt phase discontinuity for shaping the phase front of light have been reported (15). Arbitrary phase profiles along the interface can be realized by varying the geometry of each individual plasmonic antenna, which partially converts the linearly polarized light into its cross polarization with a discontinuity in phase. On the basis of this principle, anomalous reflection and refraction (16-19), broadband polarization conversion (20), and flat lenses have been realized (21-23). The ability to manipulate polarization states or phase distributions has been greatly improved through the use of metasurfaces (24,25). However, polarization states and phase distributions could not be simultaneously fully controlled in these works. The tuning and tailoring of the polarization states and phase distributions of light as two separate function parameters are necessary for the design of new-style optical devices.

Here, we theoretically predict and experimentally verify that the proposed dual-metasurfaces can provide an excellent means to simultaneously manipulate the phase and polarization of transmission light at the nanoscale. We designed a dual-metasurface to introduce a constant gradient of phase discontinuity along the interface and generated near-perfect anomalous refraction of transmission light with controllable polarization. Meanwhile, we obtained arbitrary vector optical fields by arranging the dual-layer rectangular nanoapertures. A radially polarized beam, as an example, was generated in the experiments, thereby demonstrating the power of complex beams as a design tool. We believe that the dual-metasurfaces may pave the way to significant advances in a wide range of plasmonic applications.

Figures 1A and B illustrate the top and lateral views of a fundamental unit consisting of dual-layer rectangular nanoapertures. The calculated results demonstrate that the surface plasmon polaritons (SPPs) can be excited at the metal dielectric interface and transmit through the nanoapertures to the other side of the nanoaperture film when the normally incident light is polarized perpendicular to the long axis of the rectangular nanoaperture. The standing wave of SPPs is formed in the gap of the dual-layer nanoapertures film, which is constituted by the metal-insulator-metal (MIM) waveguide. The phase and amplitude of transmission light are sensitive to the geometrical parameters and positions of the dual-layer nanoapertures forming the standing wave (26). If the dual-layer nanoapertures are aligned with each other ($S = 0$), a large phase shift of transmission light can be achieved by adjusting the length $L$ of the dual-layer nanoapertures (Fig. 1C). In addition, the ideal amplitude of transmission can be maintained while widely changing the length $L$ of dual-layer nanoapertures because of the excitation of the surface waves and localized resonance (27). The range of the controllable phase can be further extended by adjusting the lateral shift of the bottom nanoapertures. The standing wave equation between two layers can be described as $2A\sin(\omega_{spp}t)\sin(k_{spp}x)$, where $\omega_{spp}$ and $k_{spp}$ denote the angular frequency and wave number of the SPPs, respectively. The SPPs' phase will abruptly change by π at the node of the standing wave. Strong electric resonance of out-coupling SPPs can occur when the bottom nanoapertures are located at the node of the standing wave $S = \frac{P}{2}$ (28). The resonance direction of the out-coupling SPP is opposite that of $S = 0$. Therefore, the phase of the transmitted field will change by approximately π compared with that of $S = 0$. Figures 1D and E illustrate that the transmitted fields undergo a phase shift of approximately π from $S = 0$ to $S = \frac{P}{2}$, which is consistent with the previously discussed theoretical analysis. The dual-layer nanoapertures offer more degrees of freedom to control the phase of transmission light through adjustment of the parameters $L$ and $S$. Moreover, the SPPs can only be excited by the incident light polarized perpendicular to the long axis of the rectangular nanoaperture, leading to extraordinary light transmission with the same polarization. The orientation of the rectangular nanoapertures determines the polarization direction of the transmission light. On the basis of these properties of the dual-layer nanoapertures, numerous fascinating phenomena and applications can be realized, which may open the possibility of simultaneously manipulating the phase and polarization of transmission light at the nanoscale.

According to the theory of dual-layer nanoapertures, the dual-metasurface with a constant phase gradient along the $x$-axis has been designed, as illustrated in Fig. 2A.

The supercell consists of six nanostructures based on dual-layer nanoapertures, which have a similar transmission amplitude and a π/3 phase increase from left to right at a selected wavelength of 900 nm polarized in the *y*-direction, as illustrated in Figs. 2B and C. The generalized Snell's law for anomalous refraction is expressed as $n_t \sin\theta_t - n_i \sin\theta_i = \frac{\lambda_0}{2\pi}\frac{d\Phi}{dx}$ (15). Because a constant gradient of phase $\frac{d\Phi}{dx}$ exists, the phenomenon of anomalous refraction can be achieved. Because of the selective transmission of the rectangular nanoapertures, the polarization of the anomalous refractive light through the proposed dual-metasurface can be controlled. In the case of normally incident light with circular polarization, only the e-field component perpendicular to the long axis of the rectangular nanoaperture can transmit the dual-metasurface (as illustrated in Fig. 3A). Figures 3B and C show the distribution of the calculated $E_x$ and $E_y$ fields in the *x-z* plane under illumination by circular polarized incident light. The wavefront of the plane wave with *y*-polarization is generated, and a distinct deflection in the far field is formed, which is consistent with the generalized Snell's law. However, the e-field component with *x*-polarization cannot transmit the dual-metasurface. Many metasurfaces proposed in previous works can only control the phase of the portion of the transmission light with polarization orthogonal to the incidence, which leads to anomalous and ordinary refraction existing simultaneously. In contrast, our proposed dual-metasurface allows the phase of the entire transmission light with polarization parallel to the incidence to be manipulated (as illustrated in Fig. 3A). In addition, the polarization of the transmission light can depend on the orientation of the nanostructure. Therefore, simultaneous manipulation of the phase and polarization of transmission light can be realized, which further extends the ability to control light compared with the methods described in previous works (15-25). The anomalous refraction performance of the dual-metasurface can also be achieved over a broad incident angle range. The electric field distributions at three different oblique incident angles of -15°, 0°, and 15° are simulated in Figs. 3D to F through illumination by the linearly incident light polarized along the $y$-direction. The corresponding refraction angles are 45.26°, 27.73°, and 11.12°, respectively, which agree well with the calculated deflected angles of 45.70°, 27.19°, and 11.42° given by the equation of the generalized Snell's law. For the 0° incident case, the conversion efficiency from the incident wave to anomalous refraction was approximately 30%.

In the following section, we experimentally prove our theoretical design by demonstrating anomalous refraction phenomena. A representative sample with dual-layer nanoapertures is presented in Fig. 4A. Sputtering deposition, electron-beam lithography, and reactive-ion etching were used to fabricate the samples (28). The images demonstrate that the size and positioning of the nanostructure faithfully reproduce the dimensions of the design. To characterize the refraction properties of the sample, we performed far-field measurements based on the experimental setup illustrated in Fig. S1. We experimentally investigated the normalized far-field intensity at various observation angles under normal illumination by circularly polarized incident light with a wavelength of 900 nm (Fig. 4B). The experiment results indicate that only one peak of far-field intensity (the refraction angle) is observed at 27°, which implies that all of the transmission light was deflected and that the ordinary refraction light disappeared. By analyzing the polarization of the refraction light, we also observed that the transmission light is standard linear polarized light perpendicular to the long axis of the rectangular nanoaperture. The experimental results verify the ability of the sample to manipulate the propagation direction and the polarization of transmission light and agree well with the simulated results and theoretical predictions of the generalized Snell's law. In addition, the anomalous refraction phenomena under various incident angles and wavelengths were investigated. Figure 4C presents a plot of the observation angles as a function of the incident angle at a fixed wavelength of 900 nm; the results coincide well with the simulated and theoretical results (the black lines and circles). Because of the constant gradient of phase discontinuity on the surface, the incident and refraction light can be located on the same side of the surface normal in the gray region of Fig. 4C, which is an unusual "negative" refraction phenomenon. This anomalous refraction phenomenon can also be obtained within a wide wavelength range of 800 to 960 nm for normally linearly polarized incident light, as demonstrated in Fig. 4D. The refraction angle was observed to increase when the incident wavelength was increased. In addition, the large transmission amplitude was maintained in this wavelength range (Fig. S5). The measurement results are in very good agreement with the theoretical predictions. The slight differences between the measured and theoretical results are due to the inevitable structural imperfections of the fabricated sample and the inaccuracy of the Au Drude model adopted in our simulations. These results demonstrate that the ability of the dual-metasurface to manipulate the propagation direction and polarization of transmission light is applicable to a wide range of wavelengths and oblique incident angles. The realization of anomalous refraction demonstrates the abilities of dual-metasurfaces in controlling the phase of transmission light at the nanoscale.

The dual-metasurfaces composed of rectangular nanoapertures and MIM waveguides can simultaneously manipulate the phase and polarization of light at the nanoscale, which expands the degrees of freedom in controlling light. Each nanostructure composed of the dual-metasurfaces can be treated as an independent pixel because of the weak interaction. An arbitrary spatial discontinuous phase distribution can be obtained by special arrangement of the pixels. The polarization direction of the transmission light depends on the orientation of the nanoapertures. The manipulation of the phase and polarization are independent of each other. When the orientation of the nanoaperture is maintained, the six nanoapertures shown in the rows of Fig. 5 can cover the entire phase range of 0-2π, and maintain the polarization direction of the transmission light. An arbitrary polarization direction can also be obtained through rotation of the rectangular nanoapertures under illumination by circularly polarized light. In the process of polarization conversion from a circularly polarized beam to a linearly polarized beam, an additional phase (Pancharatnam–Berry phase) $e^{i\theta}$ is introduced in the transmission light, where $\theta$ is the angle of rotation (29,30). In addition, the six nanoapertures shown in the columns of Fig. 5 have a constant phase difference between each other, which can be used to generate a conjugate phase $e^{-i\theta}$. Therefore, the combination of the additional phase resulting from the rotation of the rectangular nanoaperture and the constant phase between the six nanoapertures can result in an identical phase of transmission light with different polarization directions. Therefore, 36 combinations between the phase and polarization direction can be obtained by the 36 nanostructures, as illustrated in Fig. 5. An arbitrary spatial field distribution of the optical phase and polarization direction within these 36 cases can be generated through simple arrangement of the nanostructures.

The metasurfaces proposed by the authors of previous works have been used to generate an optical vortex beam (3,31) and a twisted vector field (32). However, these metasurfaces can only individually control the light phase or the light polarization to generate some specific fields, which may limit their practical applications. When the polarization states and phase distributions are simultaneously fully controlled, arbitrary vector optical fields can be generated by special arrangement the pixels of the nanoapertures. We representatively designed and experimentally demonstrated a dual-metasurface to generate a radially polarized beam upon illumination by normally incident circularly polarized light. Figures 6A and B present SEM micrographs of the dual-metasurface. The rectangular nanoapertures in the top metal film have structural imperfections in the fabricated sample, which may slightly affect the generation of the radially polarized beam. Six types of nanoapertures, indicated by the red dashed line in Fig. 5, are arranged at spatially varying orientation angles, which generate the radial polarization direction because of the selective transmission of the rectangular nanoapertures. In addition, the phase difference between the six nanoapertures can offset the spatially varying phase distributions $e^{i\theta}$ caused by the conversion of polarization states. Therefore, a radially polarized beam can be obtained when the normally incident circularly polarized light transmits the designed dual-metasurface (Fig. 6C). The generated radially polarized beam has a characteristic doughnut intensity profile in the cross-section, as viewed by a mid-infrared camera (Fig. 6D). By rotating the linear polarizer in front of the beam, we demonstrated that its polarization is oriented radially, which is consistent with typical radially polarized beams. Therefore, the generation of vector optical fields by phase and polarization discontinuities demonstrates the versatility of the proposed dual-metasurfaces.

In summary, the proposed dual-metasurfaces first achieve full control of the phase and polarization direction of light at the nanoscale through a simple arrangement of the dual-layer rectangular nanoapertures. We theoretically designed and experimentally generated near-perfect anomalous refraction, whose polarization was controllable. Arbitrary vector optical fields were obtained using the dual-metasurfaces, and a radially polarized beam, as an example, was generated in the experiment, demonstrating the power of complex beams as a design tool. Because of the high degree of freedom in the manipulation of the phase and polarization, the dual-metasurfaces are expected to profoundly affect a wide range of optical and plasmonic applications. For instance, an appropriately constructed device can serve as a high-performance spatial light modulator, which can easily generate complex vector optical fields at the nanoscale. Two or more layers of such dual-metasurfaces can also be employed to slow the light by creating additional phase shifts in the propagation direction (24). In addition, an arbitrary distribution of the relative permittivity and permeability at an interface can be obtained through manipulation of the transmission light phase using dual-metasurfaces, which may facilitate the experimental realization of "computational metamaterials" (33). We also believe that dual-metasurfaces are potentially useful in other applications such as polarization converters, invisibility cloaking, and nanostructured holograms.

**Acknowledgments:** This work was supported by the National Basic Research Program (973 Program) of China (2012CB921900 and 2009CB930502), the Chinese National Key Basic Research Special Fund (2011CB922003), the Natural Science Foundation of China (61378006, 11304163, 11174362, 91023041, 91323304 and 61390503), the Program for New Century Excellent Talents in University (NCET-13-0294), the Knowledge Innovation Project of CAS (Grand No. KJCX2-EW-W02), the Natural Science Foundation of Tianjin (13JCQNJC01900), and the 111 project (B07013).

**Supplementary Materials**

Materials and Methods

Figs. S1 to S8

References (*34–49*)

Movies S1 and S2

# Figures

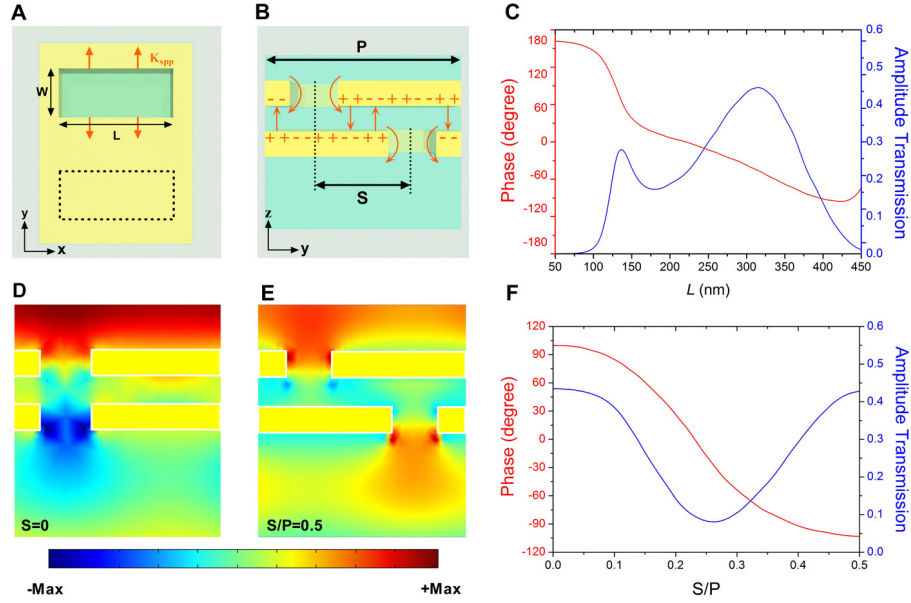

**Fig. 1.** Schematic illustration of the designed unit of a dual-layer rectangular nanoaperture film for (**A**) *x-y* plane and (**B**) *y-z* plane cross-sections. The two metal films of the MIM waveguide have the same thickness of 70 nm and are separated by a distance of 70 nm; the films are wholly embedded in silicon dioxide. The rectangular nanoapertures in the two layers have the same dimensions. The lateral shift along the *y*-axis is defined as parameter *S*. The SPPs can be excited by linear light at 900 nm polarized perpendicular to the long axis of the rectangular nanoaperture. The excited SPPs can transmit through the rectangular nanoaperture film and form the extraordinary optical transmission. Meanwhile, the standing wave can be formed in the MIM waveguide. The plus (minus) signs and orange arrows in (B) indicate the e-field oscillation of the SPPs. (**C**) The calculated phase (red line) and amplitude (blue line) of the transmission light from the dual-layer nanoaperture film for various values of *L*, where the width of the nanoaperture *W* is 130 nm and the lateral shift *S* is 0. (**D**

and **E**) Simulated $E_y$ field patterns for the aligned and dislocated dual-layer rectangular nanoaperture film in the *y-z* plane. (**F**) Calculated phase (red line) and amplitude (blue line) of the transmission light as a function of lateral shift. The length *L* and width *W* of the rectangular nanoaperture in (D-F) are 300 nm and 130 nm, respectively. The period distances in the *x* and *y* directions are 400 nm and 530 nm, respectively.

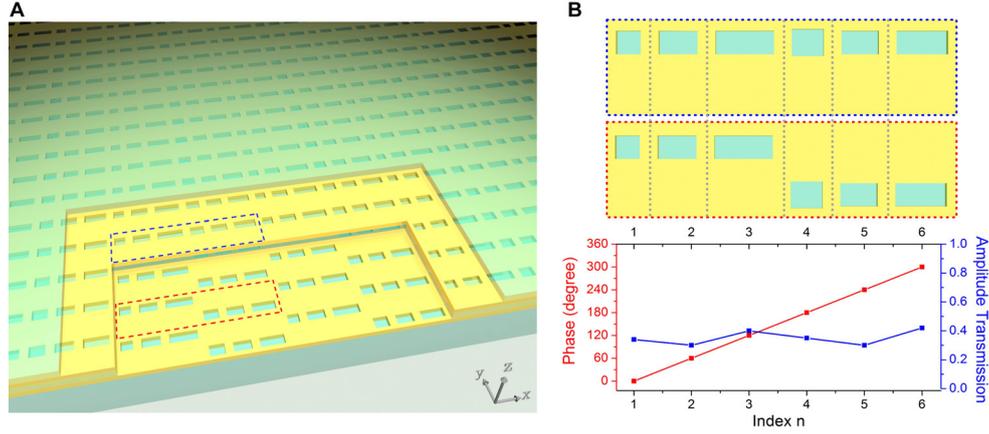

**Fig. 2.** (**A**) Three-dimensional schematic view of the designed dual-metasurfaces. Some parts of the silicon dioxide and Au film are uncovered to reveal the nanostructures in the Au film. (**B**) Upper panel: Schematics of the upper and bottom supercell of the sample (corresponding to the regions surrounded by the dashed lines in (A)). The optimized geometrical parameters of the nanoapertures are $L$ = 140, 220, 330, 180, 210, and 290 nm with corresponding $W$ = 130, 130, 130, 150, 130, and 130 nm, respectively. The distance between adjacent nanoapertures along the $x$-axis is fixed at 100 nm. The lateral shift $S$ for the first and second three nanoapertures are 0 and 265 nm, respectively. The six nanoapertures are treated as a supercell element embedded in the silicon substrate with periods $P_x$ = 1970 nm and $P_y$ = 530 nm, thereby forming a linear phase profile with a $2\pi$ range at the interface. Lower panel: The transmission phase and amplitude of each structural unit within a supercell, calculated under $y$-polarized incidence at 900 nm.

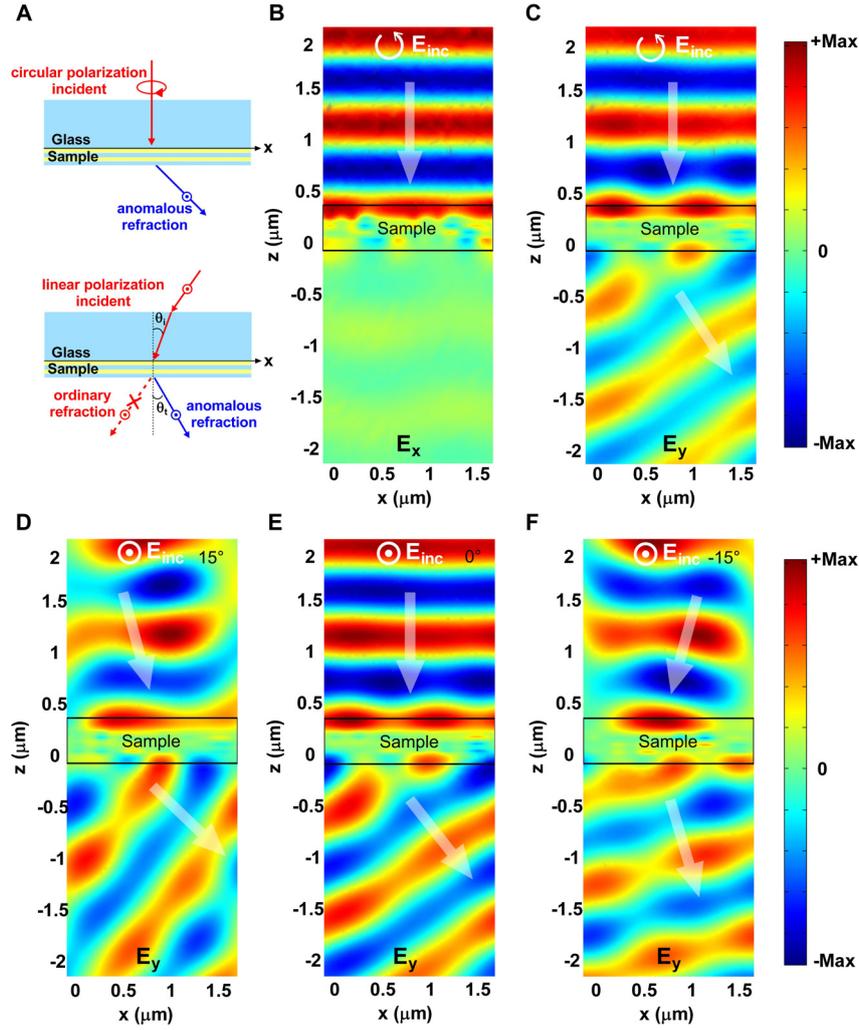

**Fig. 3.** (**A**) Schematic illustration of anomalous refraction by dual-metasurface when illuminated with circularly and linearly polarized incident light. Only the e-field component perpendicular to the long axis of the rectangular nanoaperture can transmit the dual-metasurface, where ordinary refraction does not exist. (**B** and **C**) Simulated $E_x$ and $E_y$ field patterns in the *x-z* plane, excited by normally circularly polarized incident light. (**D-F**) Simulated $E_y$ field patterns in the *x-z* plane with *y*-polarized incident light for incidence angles θ =15°, 0°, and -15°. The incident wavelength was fixed at 900 nm.

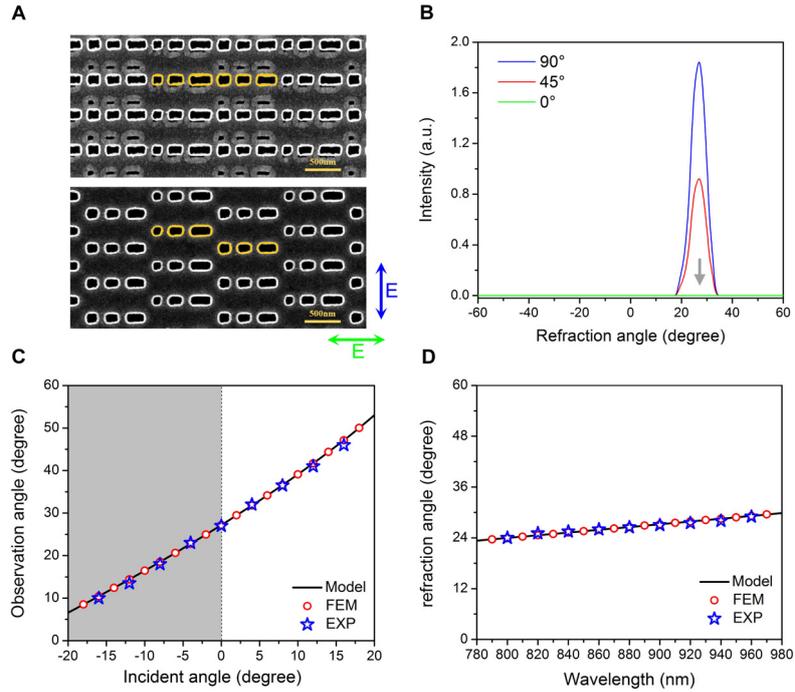

**Fig. 4.** (**A**) Scanning electron microscopy (SEM) images of dual-metasurface fabricated on a silicon dioxide wafer. The super cell in each layer is highlighted in yellow. (**B**) Measured far-field intensity at various observation angles for circularly polarized excitations. An analyzer was inserted into the optical path to analyze the polarization status of the transmission light. The green and blue double arrows in (A) indicate the polarization status corresponding to 0° and 90° in (B). The gray arrow indicates the calculated angle of anomalous refraction based on the generalized Snell's law. (**C**) The experimental results (stars) of anomalous refraction at various incident angles under *y*-polarized incidence at 900 nm. The shaded region represents "negative" refraction for the transmission light. (**D**) Experimental results (stars) of anomalous refraction at various incident wavelengths under normally *y*-polarized incidence. The circles and solid lines in (C) and (D) are the corresponding finite element method (FEM) simulations and theoretical calculations performed using the generalized Snell's law for refraction.

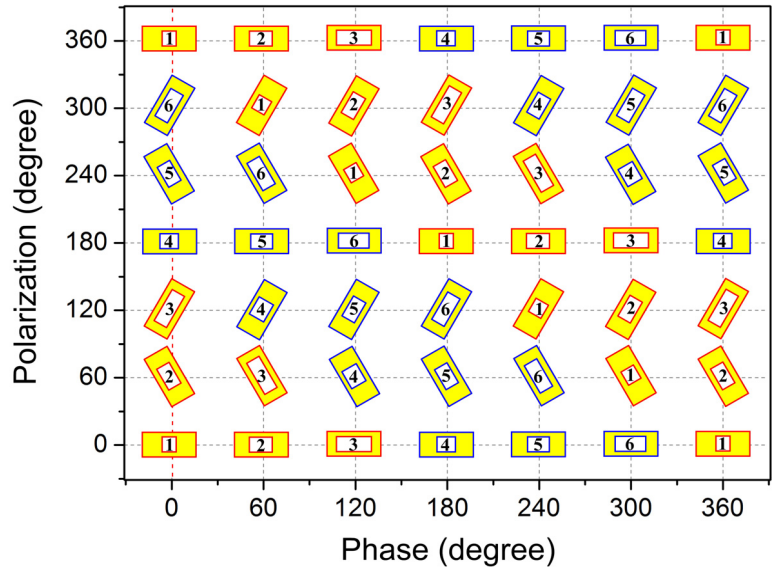

**Fig. 5.** The 36 combinations of the polarization and phase of transmission light by dual-layer nanoapertures with various dimensions and orientations. The red (blue) frames indicate the alignment (dislocation) of the dual-layer nanoapertures. The dual-layer nanoapertures have the same dimensions and lateral shift as the units in Fig. 2B. The numbers in the dual-layer nanoapertures are consistent with those in Fig. 2B.

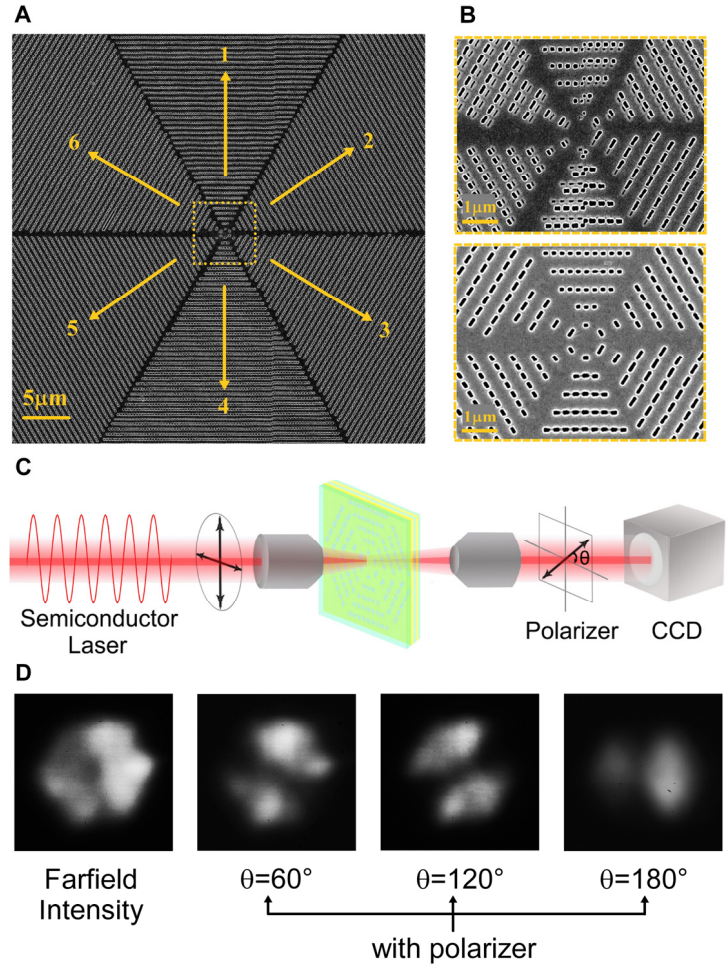

**Fig. 6.** (**A**) SEM image of the dual-metasurface, which was designed to generate a radially polarized beam. The dual-metasurface consists of six regions. Each region is filled with one type of nanoaperture, which is indicated by the red dashed line in Fig. 5. The yellow arrows represent the designed distribution of the polarization direction. The numbers in the dual-layer nanoapertures are consistent with those in Fig. 2B. (**B**) Magnified view of the center part of (A) for the upper and bottom layer, respectively. (**C**) The experimental setup for generating and detecting the radially polarized beam. (**D**) Measured far-field intensity distributions, without and with a polarizer, where the polarizer was oriented at various angles in front of the CCD camera. These intensity patterns demonstrate that the transmission beam is radially polarized.

Supporting Online Material for

# Full control of polarization states and phase distributions of light with dual-metasurfaces

**This PDF file includes:**

Materials and Methods

Figs. S1 to S8

References (*34–49*)

**Other Supplementary Material for this manuscript includes the following:**

Movies S1 and S2

**Materials and Methods**

Numerical simulations

All of our numerical simulations were performed using the finite element method (FEM)-based commercial software COMSOL Multiphysics. The permittivity of gold is described by the Drude model (34), and the dielectric constant of silicon dioxide is 2.25. For the simulation of anomalous refraction, the supercell shown in Fig. 2 has periodic boundary conditions in the $x$ and $y$ planes, and the waveguide ports boundary conditions on the other boundaries. For the simulation of the generation of the radially and angularly polarized beam, the waveguide ports boundary condition was also employed as the incident light source, and perfectly matched layers were placed around the simulation domain to completely absorb the waves leaving the simulation domain.

Sample fabrication

Sputtering deposition, electron-beam lithography, and reactive-ion etching were used to fabricate the metallic structures. The samples shown in Fig. 4 for the anomalous refraction and in Fig. 6 for the generation of the radially polarized beam were fabricated using the following steps. Au was deposited onto the bare glass substrate to a thickness of 70 nm using a radio-frequency magnetic sputtering system, and a 200-nm-thick PMMA resist was subsequently spin-coated onto the sample; the sample was then subjected to bakeout at 180°C on a hotplate for 2 min. The pattern was exposed using an electron-beam lithography system (Raith150) at 10 keV. After exposure, the sample was developed in MIBK:IPA (1:3) for 40 s and IPA for 30 s and then blown dry using pure nitrogen. The pattern was transferred onto the Au layer by a reactive-ion etching system using Ar gas. After the PMMA resist was removed with acetone, a 70-nm-thick $SiO_2$ layer and a 70-nm-thick Au layer were deposited onto the sample using plasma enhanced chemical vapor deposition (PECVD) and a radio-frequency magnetic sputtering system, respectively. The pattern on the second Au layer was prepared by repeating the exposure, etching, and resist-removal processes. Finally, a 70-nm-thick $SiO_2$ layer was deposited on top of the sample using PECVD.

Measurement setup

A schematic of the measurement setup for the anomalous refraction is presented in Fig. S1. A mode-locked Ti:sapphire oscillator provided near-infrared femtosecond laser pulses with a central wavelength of 800 nm, a pulse duration of ~120 fs, and a repetition rate of 1 kHz. The wavelength of light from the laser was tuned over a wide range using an automated optical parametric amplifier (OPAS Prime). A polarizer and a QWP were combined to generate the incident circularly polarized light, which was focused on the sample with a 20×/0.40 NIR microscope objective. The transmission through the sample was collected using a lens. For the refractive angle detection, we used two concentric rotary systems to achieve independent rotation of the sample orientation and the detector angle. The resolution of the rotation angle was 0.02°. The lens, analyzer, and power meter in the rotary system were used to measure the intensity of light. The analyzer was inserted into the optical path to analyze the polarization status of the transmission light. The other rotary stage was used to adjust the orientation of the sample for oblique incident while the position of the sample was maintained at the center of the stage. This setup allowed us to verify the anomalous refraction phenomenon generated by the sample. All the optical elements, including the microscope objective, lens, QWP, polarizer, and detector, were operated in the broadband range.

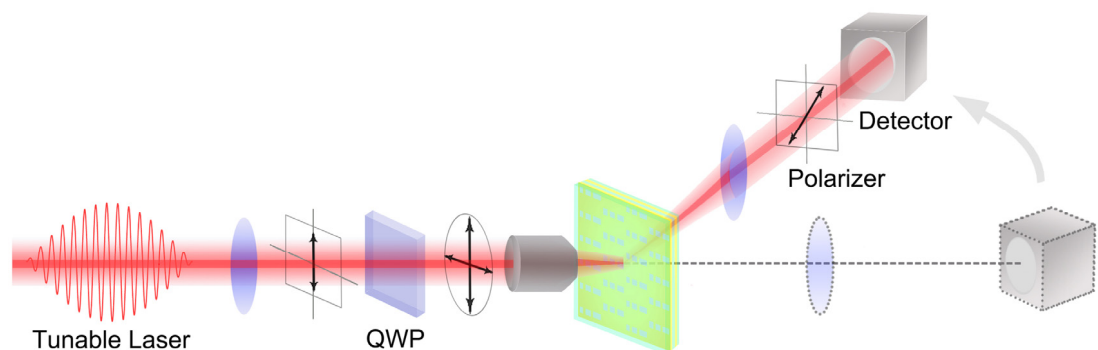

**Fig. S1.** A schematic of the measurement setup for the detection of anomalous refraction. QWP: quarter-wave plate.

**SOM Text**

Excitation of SPP standing waves in an MIM waveguide

The excitation of surface plasmon polaritons (SPPs) in an array of subwavelength holes has recently received substantial attention (35-37). The controllable properties of SPPs that propagate in the plane of the metal film have led to numerous advances in the manipulation of light at the nanoscale (38-41). Rectangular nanoapertures can act as a point dipole source of SPPs and play an important role in a series of related research, such as the focusing of SPPs (38) and the control of directional coupling (41). Metal-insulator-metal (MIM) waveguides are promising in the manipulation of the propagation of SPPs because their outstanding properties of high spatial field confinement and micrometer-range propagation lengths (42-44). Therefore, an MIM waveguide with a slit or rectangular nanoaperture in the top metal film can combine the excitation sources of SPPs and plasmonic waveguides. When a light polarized perpendicular to the long axis of a rectangular nanoaperture is normally incident to the metal film, SPP beams with a $\pi$ phase difference can be excited at two sides of the rectangular nanoaperture and propagate in opposite directions on the metal-dielectric interface, as illustrated in Fig. S2A. Some of the excited SPPs can transmit through the nanoapertures to the other side of the film and be coupled in the MIM waveguide (these SPPs are termed gap plasmons), and the standard interference pattern of an SPP standing wave can be obtained in the MIM waveguide. Figure S2B shows the electric field distribution of the SPP standing wave in the MIM waveguide. The wavenumber of the SPP in the MIM waveguide is $k_{spp} = \dfrac{2\pi n_{eff}}{\lambda_{inc}}$, and $n_{eff}$ is the effective refractive index of the MIM waveguide, which can be calculated from the following formulas (45):

$$\tanh \dfrac{dU}{2} = -\dfrac{\epsilon_i W}{\epsilon_m U}, \tag{2}$$

$$U = k_0 (n_{\text{eff}}^2 - \epsilon_i)^{1/2}, \tag{3}$$

$$W = k_0 (n_{\text{eff}}^2 - \epsilon_m)^{1/2}, \tag{4}$$

where $d$ is the distance between the two metal films, and $\epsilon_i$ and $\epsilon_m$ are the permittivities of the insulator and metal, respectively. In Fig. S2, the calculated $n_{eff}$ is approximately 2.

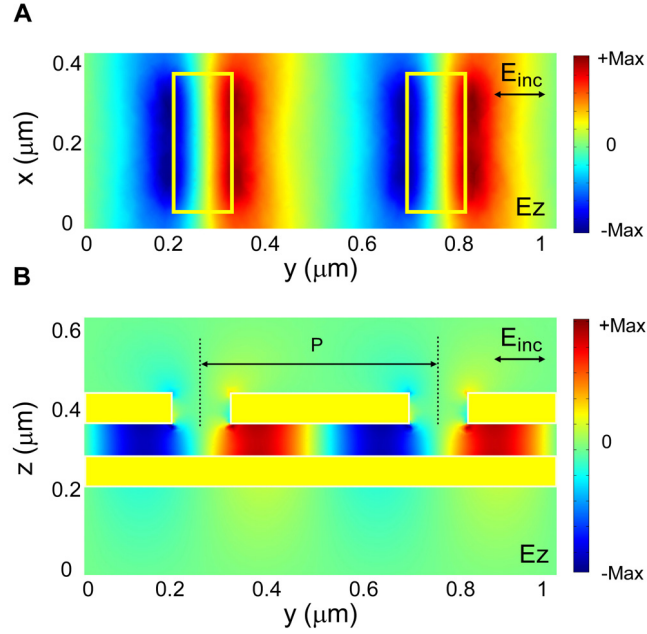

**Fig. S2.** Simulated $E_z$ field patterns of an MIM waveguide with periodic rectangular nanoapertures in the top metal film in (**A**) the *x-y* plane and (**B**) the *y-z* plane. The rectangular nanoapertures are the same as those in Fig. 1. The incident light is linearly polarized along the *y*-axis at a wavelength of 900 nm.

Out-coupling of SPPs in the MIM waveguide

If the same rectangular nanoapertures are also inserted in the bottom metal layer, the SPPs propagating in the MIM waveguide can out-couple to free-space radiation, which generates the transmission light. The positions of the bottom rectangular nanoapertures will greatly affect the amplitude and phase of the transmission light. If the nanoapertures are deposited at the antinodes of the standing wave, the out-coupling SPPs cannot form a distinct electric resonance with the same phase of SPP resonance as the bottom nanoapertures, resulting in low transmission. If the nanoapertures are deposited at the nodes of the standing wave, strong electric resonance and a large conversion from the SPPs to transmission light can be achieved because the out-coupling SPPs has a resonance with the opposite phase at the bottom nanoaperture. The phase difference of the transmission light will change distinctly upon adjustment of the position of the bottom nanoaperture, which can be used to expand the tunable range of the phase. Figure S3 shows the $E_z$ field distributions of dual-layer rectangular nanoaperture film for $S = 0$ and $S = \frac{P}{2}$. The resonance direction of out-coupling SPP is apparently opposite in the two cases. Therefore, the phase of the transmitted field will approximately change $\pi$ from $S = 0$ to $S = \frac{P}{2}$.

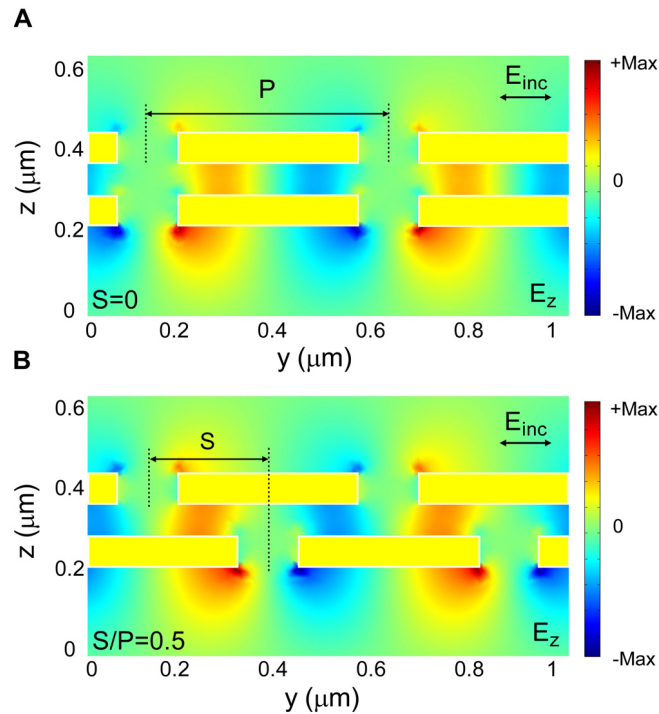

**Fig. S3.** Simulated $E_z$ field patterns of a dual-layer rectangular nanoaperture film in the *y-z* plane for (**A**) $S = 0$ and (**B**) $S = \frac{P}{2}$. The incident light is linearly polarized along the *y*-axis at a wavelength of 900 nm. The position of $S = \frac{P}{2}$ corresponds exactly with the node of the standing wave.

Selective transmission of dual-metasurfaces

The rectangular nanoapertures in the metal film can selectively transmit light polarized perpendicular to their long axis because of the excitation of SPPs (46). Therefore, the polarization of the transmitted field completely depends on the orientation of the rectangular nanoaperture under illumination by circularly polarized light. Dual-metasurfaces consisting of dual-layer rectangular nanoaperture film have the same characteristics as a single-layer rectangular nanoaperture film. Figure S4 presents a plot of the simulated *x* and *y* components of the transmission amplitude for the aligned (offset) dual-metasurface as a function of *L* and *W* with normally circularly polarized incident light with a wavelength of 900 nm. This figure demonstrates that the six nanostructures proposed in this paper can only transmit light with a *y* e-field component, which is vertical to the long axis of the rectangular nanoaperture. Because of this extraordinary optical property of dual-metasurfaces, the polarization direction of transmission light can be controlled through rotation of the nanostructure by a desired angle.

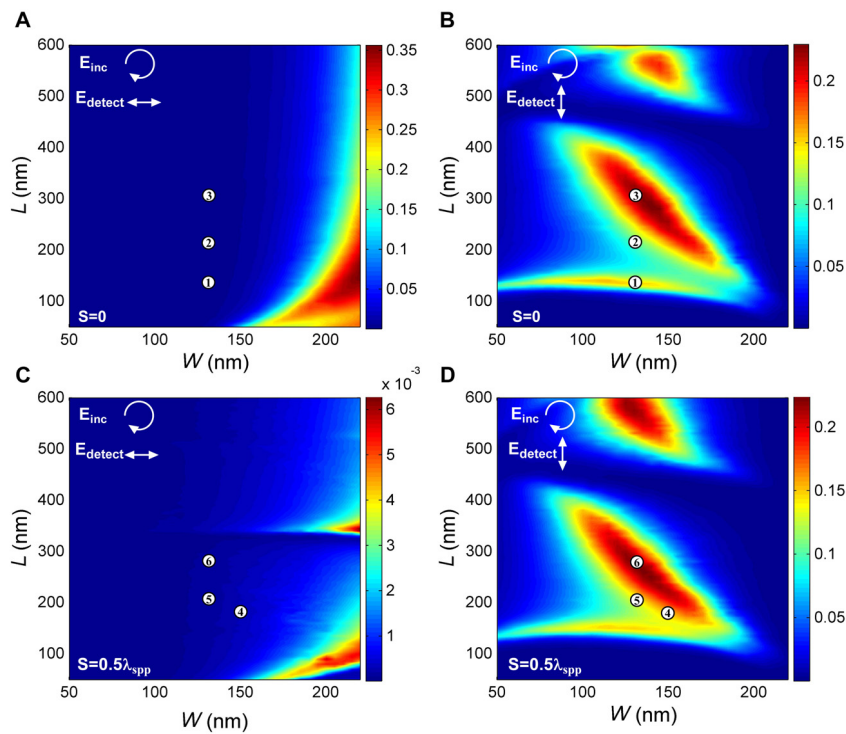

**Fig. S4.** Simulated *x*-component (**A** and **C**) and *y*-component (**B** and **D**) transmission amplitude of the aligned (**A** and **B**) and dislocated (**C** and **D**) dual-layer rectangular nanoapertures at various *L* and *W* with normally circularly polarized incident light at 900 nm. The lateral shift was fixed at 265 nm. The circles with numbers indicate the values of *L* and *W* used in the experiments, which correspond to the nanostructures defined in Fig. 2**B**.

Broadband functionality of anomalous refraction

The generation of anomalous refraction occurs because of the capabilities of dual-metasurfaces in the manipulation of the optical phase at the nanoscale; this behavior can be understood using the generalized Snell's law (15):

$$\theta_t = \sin^{-1}(\frac{\lambda_0}{2\pi n_t}\frac{d\Phi}{dx} + \frac{n_i}{n_t}\sin\theta_i), \quad (5)$$

where $\theta_i$ is the incident angle and $\frac{d\Phi}{dx}$ is the gradient of the phase. Figure S5 shows the measured normalized scattered electric field intensity at various observation angles with various incident wavelengths under the normal illumination by *y*-polarized light. The phenomenon of anomalous refraction can be obtained in a large wavelength range, and the refraction angle increases with increasing incident wavelength, which is consistent with the theoretical predictions of the generalized Snell's law. Meanwhile, the large transmission amplitude can also be maintained in a broadband wavelength range. The broadband functionality implies that the phase increments between the nearest neighbors are not rigorously necessary; however, the overall trends of the phase shift and the uniform transmission amplitude along the interface are indeed indispensable.

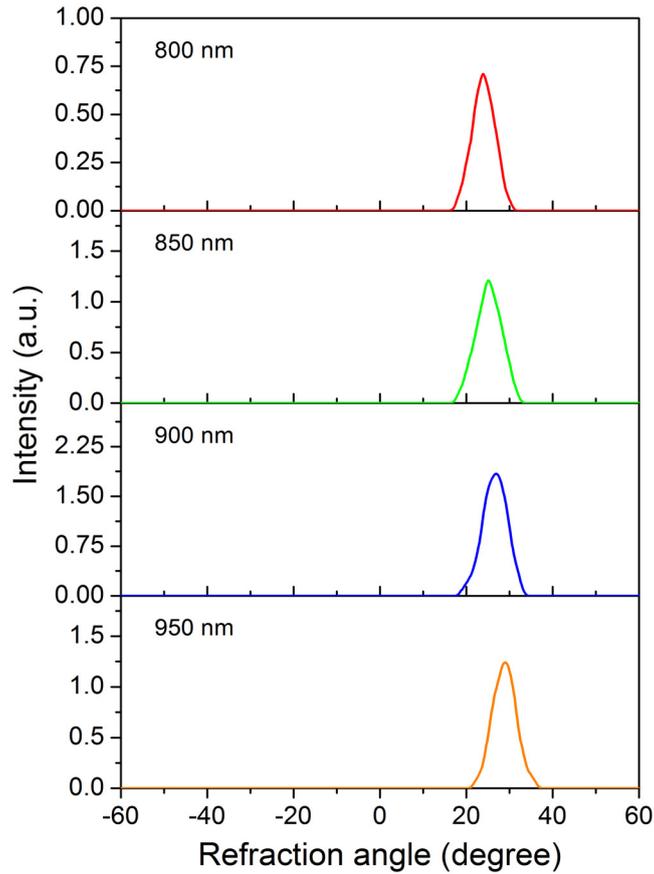

**Fig. S5.** Measured far-field intensity at various observation angles for circularly polarized excitations with various wavelengths.

Simulation of the generation of vector beams

Radially polarized beams can be obtained using numerous methods, including the superposition of two cross-polarized Hermite−Gauss modes (47) or the selective coupling of free space light to specific transverse modes (48,49). These methods all require a complex combination of multiple optical components. By simultaneously fully controlling the polarization and phase at the nanoscale, diverse vector optical fields can be generated through the design of appropriate dual-metasurfaces. We have experimentally demonstrated the generation of a radially polarized beam in Fig. 6. By imaging with a charge-coupled device camera (CCD), we proved that the distribution of phase and the polarization of the transmission beam satisfies the standard of a radially polarized beam. We also simulated the generation of the radially and angular polarized beams in Fig. S6. Figures S6A and C show schematics of the simulated modes, which are divided into six regions corresponding to six types of nanostructures. The spatially varying orientation of rectangular nanoapertures in the different regions results in the radial or annular distribution of linear polarization. Figures S6B and D show the e-field distributions of the transmission beam.

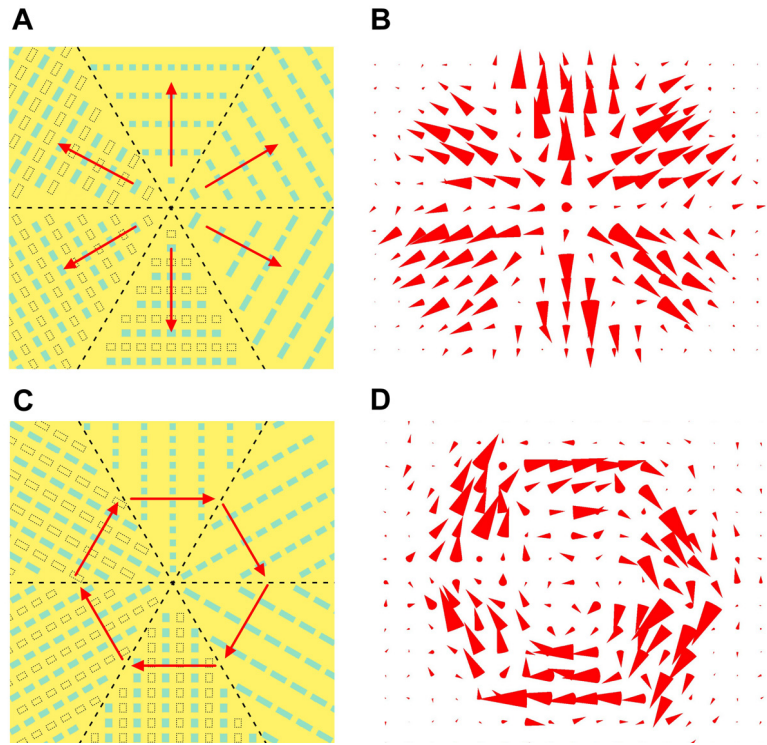

**Fig. S6** (**A** and **C**) A schematic of the simulated modes for generating the radially and angularly polarized beam, respectively. The arrows indicate the desired distribution of the polarization direction. (**B** and **D**) The e-field distribution of the transmission beam on the cross-section in the propagation direction. The blue and yellow rectangles represent the positions of the nanoapertures in the top and bottom layer, respectively.

**Movies S1 and S2**

Movies S1 and S2 show the e-field oscillation of transmission beams on the cross-section in the propagation direction, which further demonstrates the generation of the radially and angular vector beams by the designed dual-metasurfaces.